\documentclass[aps,prl,twocolumn,groupedaddress,pdfLaTeX]{revtex4-2}

\usepackage{amsmath}
\usepackage{bm}
\usepackage{graphicx}
\usepackage{hyperref}
\hypersetup{colorlinks=true,citecolor=magenta,linkcolor=magenta,urlcolor=magenta}

\begin{document}


\title{Universal Digitized Counterdiabatic Driving}


\author{Takuya Hatomura}
\email[]{takuya.hatomura@ntt.com}
\affiliation{Basic Research Laboratories \& NTT Research Center for Theoretical Quantum Information, NTT, Inc., Kanagawa 243-0198, Japan}


\date{\today}

\begin{abstract}
Counterdiabatic driving realizes parameter displacement of an energy eigenstate of a given parametrized Hamiltonian using the adiabatic gauge potential. 
In this paper, we propose a universal method of digitized counterdiabatic driving, constructing the adiabatic gauge potential in a digital way with the idea of universal counterdiabatic driving. 
This method has three advantages over existing universal counterdiabatic driving and/or digitized counterdiabatic driving: it does not introduce any many-body and/or nonlocal interactions to an original target Hamiltonian; it can incorporate infinite nested commutators, which constitute the adiabatic gauge potential; and it gives explicit expression of rotation angles for digital implementation. 
We show the consistency of our method to the exact theory in an analytical way and the effectiveness of our method with the aid of numerical simulations. 
\end{abstract}


\maketitle

{\it Introduction.---}
Adiabatic control is a fundamental means of tailoring quantum states with slow parameter changes~\cite{Born1928,Kato1950}. 
The adiabatic gauge potential (AGP) plays an important role in the theory of adiabatic control~\cite{Kolodrubetz2017}. 
It has various applications: it is a mathematical tool to prove the adiabatic theorem~\cite{Kato1950}; it is used to analyze near-adiabatic dynamics~\cite{Rigolin2008,Suzuki2020,Hatomura2020}; and it can be a probe to explore quantum phase transitions~\cite{You2007,Zanardi2007} and quantum chaos~\cite{Pandey2020,Lim2024}. 
AGP is also central to the theory of shortcuts to adiabaticity, which enables to speed up adiabatic control~\cite{Torrontegui2013,Guery-Odelin2019,Hatomura2024}. 
Shortcuts to adiabaticity are promising candidates for improving quantum adiabatic algorithms, e.g., quantum annealing and quantum adiabatic computation~\cite{Albash2018}, and thus have received much attention.

Counterdiabatic driving is a method of shortcuts to adiabaticity using AGP~\cite{Demirplak2003,Demirplak2008,Berry2009}. 
Originally, the calculation of AGP required knowledge of energy spectra, but the theory of variational counterdiabatic driving reformulated it as a variational minimization problem without that knowledge~\cite{Sels2017}. 
This variational approach is based on the algebraic characterization of AGP, and thus it can also be reformulated as an algebraic problem~\cite{Hatomura2021,Xie2022,Takahashi2024,Bhattacharjee2023,Ohga2025}. 
More recently, universal counterdiabatic driving was proposed, based on universal polynomial fitting of AGP~\cite{Morawetz2025,Finzgar2025}. 
These approaches theoretically enable to calculate AGP, but they require incredible computational cost because the number of relevant operators exponentially increases with system size in many-body systems.

Truncation of many-body and/or nonlocal interactions in AGP is a typical approach to avoid expensive computational cost. 
Truncated AGPs can be expressed in terms of nested commutators of a target Hamiltonian and its parameter derivative~\cite{Claeys2019}, and their coefficients can be numerically calculated with existing methods~\cite{Sels2017,Hatomura2021,Xie2022,Takahashi2024,Bhattacharjee2023,Ohga2025,Morawetz2025,Finzgar2025}. 
Moreover, such terms can be effectively implemented using Floquet engineering~\cite{Petiziol2018,Claeys2019}. 
Besides Floquet engineering, digital quantum simulation is another approach to the realization~\cite{Hegade2021,Hegade2022,Chen2022,Hatomura2023,Keever2024,Vreumingen2024,Hatomura2025,Vizzuso2025,Bhargava2026}. 
These approaches are promising, but experimental realization is still limited up to a few-order nested commutators. 
A main obstacle in Floquet engineering is the requirement of high-frequency driving, and that in digital quantum simulation is the necessity of deep circuit depth.

In this paper, we propose {\it universal digitized counterdiabatic driving}, a universal method of calculating AGP and realizing counterdiabatic driving in a digital way. 
We emphasize that it is not simple digitization of universal counterdiabatic driving, but it is a pivotal universal method of digitized counterdiabatic driving. 
Our method does not introduce any many-body and/or nonlocal interactions, which are not included in a target Hamiltonian. 
Nevertheless, we can digitally generate infinite nested commutators, which constitute AGP. 
Moreover, we can obtain explicit expression of rotation angles for digital implementation. 
In addition to these clear features, our analysis shows that the present method leads to the resolution of the previous problems mentioned above, making it promising.

%
%
{\it Theoretical background.---}
We briefly summarize the theory of counterdiabatic driving, from the definition of AGP to the idea of universal counterdiabatic driving. 
For a given parametrized Hamiltonian of a $D$-dimensional quantum system, $\hat{H}(\lambda)=\sum_{n=0}^{D-1}E_n(\lambda)|n(\lambda)\rangle\langle n(\lambda)|$, AGP is defined as
\begin{equation}
\hat{\mathcal{A}}(\lambda)=i\sum_{\substack{m,n=0 \\ (m\neq n)}}^{D-1}|m(\lambda)\rangle\langle m(\lambda)|\partial_\lambda n(\lambda)\rangle\langle n(\lambda)|. 
\label{Eq.AGP}
\end{equation}
AGP generates small displacement of an energy eigenstate in parameter space as $|n(\lambda+\delta\lambda)\rangle=\hat{U}_\mathrm{ad}(\lambda)|n(\lambda)\rangle$, where $\hat{U}_\mathrm{ad}(\lambda)$ is a unitary operator
\begin{equation}
\hat{U}_\mathrm{ad}(\lambda)=e^{-i\delta\lambda\hat{\mathcal{A}}(\lambda)},
\label{Eq.ad.trans}
\end{equation}
with infinitesimal parameter displacement $\delta\lambda$~\cite{Kolodrubetz2017}. 
Equation~(\ref{Eq.ad.trans}) is regarded as the quench limit of counterdiabatic driving. 
Our target is to construct AGP (\ref{Eq.AGP}) for realizing the unitary operator (\ref{Eq.ad.trans}) in a practical way.

It is possible to approximate AGP (\ref{Eq.AGP}) with an Hermitian operator
\begin{equation}
\hat{A}^{(d)}(\lambda)=i\sum_{l=1}^d\alpha_l(\lambda)\mathcal{L}^{2l-1}\partial_\lambda\hat{H}(\lambda),
\label{Eq.AGP.nested}
\end{equation}
where $\mathcal{L}$ is a superoperator $\mathcal{L}\bullet=[\hat{H}(\lambda),\bullet]$~\cite{Claeys2019}. 
The coefficients $\{\alpha_l(\lambda)\}$ can be determined by using the algebraic characterization of AGP~\cite{Sels2017,Hatomura2021,Xie2022,Takahashi2024,Bhattacharjee2023,Ohga2025} or polynomial fitting~\cite{Morawetz2025,Finzgar2025}. 
It can actually be the exact AGP (\ref{Eq.AGP}) when the integer $d$ is large and the nested commutators $\{\mathcal{L}^{2l-1}\partial_\lambda\hat{H}(\lambda)\}$ constitute closed Lie algebra. 
Owing to the difficulties in computational cost and experimental implementation, small $d$, usually $d=\mathcal{O}(1)$, is adopted in practice.

The difference between the Hermitian operator (\ref{Eq.AGP.nested}) and AGP (\ref{Eq.AGP}) can be evaluated as
\begin{equation}
\|\hat{\mathcal{A}}(\lambda)-\hat{A}^{(d)}(\lambda)\|^2=\int d\omega\left[\frac{1}{\omega}+\sum_{l=1}^d\alpha_l(\lambda)\omega^{2l-1}\right]^2\Phi(\omega),
\label{Eq.distance.nested}
\end{equation}
where the norm is the Hilbert-Schmidt norm $\|\bullet\|=\sqrt{\mathrm{Tr}\bullet^2}$ and $\Phi(\omega)$ is the spectral function $\Phi(\omega)=\sum_{m,n=0 (m\neq n)}^{D-1}|\langle m(\lambda)|\partial_\lambda\hat{H}(\lambda)|n(\lambda)\rangle|^2\delta(\omega-\omega_{mn})$ with an energy difference $\omega_{mn}=E_m(\lambda)-E_n(\lambda)$. 
In universal counterdiabatic driving~\cite{Morawetz2025,Finzgar2025}, we consider polynomial fitting of $(-1/\omega)$ with the appropriate coefficients $\{\alpha_l(\lambda)\}$ for minimizing the difference (\ref{Eq.distance.nested}).

%
%
{\it Method.---}
Now we propose our method, {\it universal digitized counterdiabatic driving}. 
In this method, we consider the implementation of the following digital, composite unitary operator
\begin{equation}
\hat{U}(\lambda)=\prod_{\substack{k=-K \\ (k\neq0)}}^Ke^{i\theta_k\hat{H}(\lambda)}e^{-i\frac{\phi_k}{2}\partial_\lambda\hat{H}(\lambda)}e^{-i\theta_k\hat{H}(\lambda)}, 
\label{Eq.U}
\end{equation}
with explicit rotation angles
\begin{equation}
\theta_k=\frac{k\pi}{\Omega},\quad\phi_k=-\mathrm{sgn}(k)\frac{2\delta\lambda}{\Omega}\mathrm{Si}(\theta_k\Omega). 
\label{Eq.rot.angle}
\end{equation}
Here, $\mathrm{Si}(x)=\int_0^xdy(1/y)\sin(y)$ is the sine integral and $\Omega$ is a certain cutoff. 
As discussed later, it is natural to set $\Omega=\Delta_\mathrm{max}$, where $\Delta_\mathrm{max}$ is the maximum energy gap, while it is not necessarily to set in that way.

We explain the key ideas, advantages, and possible extension of our method (details of the derivation are available at END MATTER). 
Clearly, the composite unitary operator (\ref{Eq.U}) does not have any many-body and/or nonlocal interactions, which are not included in the target Hamiltonian $\hat{H}(\lambda)$. 
The composite unitary operator (\ref{Eq.U}) can be approximated as a unitary operator with a single generator, i.e., $\hat{U}(\lambda)=e^{-i\delta\lambda\hat{V}(\lambda)}+\mathcal{O}(\{\phi_k^2\})$ with an Hermitian operator
\begin{equation}
\hat{V}(\lambda)=i\sum_{l=1}^\infty\left[\frac{(-1)^{l+1}}{(2l-1)!\delta\lambda}\sum_{k=1}^K\phi_k\theta_k^{2l-1}\right]\mathcal{L}^{2l-1}\partial_\lambda\hat{H}(\lambda),
\label{Eq.AGP.ours}
\end{equation} 
for $\theta_k=-\theta_{-k}$, $\phi_k=-\phi_{-k}$, and $|\phi_k|\ll1$. 
The operator (\ref{Eq.AGP.ours}) has similar structure to the operator (\ref{Eq.AGP.nested}), and thus it has the potential to approximate AGP (\ref{Eq.AGP}). 
Notably, the operator (\ref{Eq.AGP.ours}) consists of infinite nested commutators unlike the operator (\ref{Eq.AGP.nested}).

The difference between the operator~(\ref{Eq.AGP.ours}) and AGP (\ref{Eq.AGP}) is given by
\begin{equation}
\|\hat{\mathcal{A}}(\lambda)-\hat{V}(\lambda)\|^2=\int d\omega\left[\frac{1}{\omega}+\sum_{k=1}^K\frac{\phi_k}{\delta\lambda}\sin(\theta_k\omega)\right]^2\Phi(\omega). 
\label{Eq.distance}
\end{equation}
Remarkably, owing to the infinite nested commutators in Eq.~(\ref{Eq.AGP.ours}), we can adopt the Fourier series expansion instead of polynomial fitting unlike universal counterdiabatic driving, and it enables to obtain the explicit expression of the rotation angles. 
Indeed, the Fourier expansion of $(-1/\omega)$ gives the rotation angles (\ref{Eq.rot.angle}). 
The spectral function $\Phi(\omega)$ is nonzero only in the ranges $[-\Delta_\mathrm{max},-\Delta_\mathrm{min}]$ and $[\Delta_\mathrm{min},\Delta_\mathrm{max}]$, where $\Delta_\mathrm{min}$ is the minimum energy gap. 
This is the reason why $\Omega=\Delta_\mathrm{max}$ is natural, while we will numerically show that it is not necessarily to set in that way. 
In summary, the validity of universal digitized counterdiabatic driving, using the composite unitary operator (\ref{Eq.U}) with the rotation angles (\ref{Eq.rot.angle}), is theoretically supported.

A possible extension of our proposal is the introduction of the regularization in AGP~\cite{Claeys2019}, which results in $(1/\omega)\to\omega/(\omega^2+\eta^2)$ ($\eta^2\ll\Delta_\mathrm{min}^2$) in Eq.~(\ref{Eq.distance}). 
It resolves a certain problem associated with the divergence of $(1/\omega)$ at $\omega=0$, which we will find by numerical simulation, while $\phi_k$ is modulated as $\phi_k=-\mathrm{sgn}(k)\frac{2\delta\lambda}{\Omega}\int_0^{\Omega}[\omega/(\omega^2+\eta^2)]\sin(k\pi\omega/\Omega)d\omega$. 
Note that, when the target state is the ground state, $\Delta_\mathrm{min}$ can be regarded as the minimum energy gap for the ground state, not the minimum energy gap between all the eigenstates.

%
%
{\it Results.---}
Then, we consider two examples. 
One is a two-level system, for which we can exactly minimize the difference (\ref{Eq.distance}) with rotation angles different from Eq.~(\ref{Eq.rot.angle}). 
This example is discussed to confirm the consistency of our basic theory to the exact one. 
The other is a many-body system, for which we consider our method proposed above and confirm effectiveness of it.

First, we consider a two-level system $\hat{H}(\lambda)=h^X(\lambda)\hat{X}+h^Z(\lambda)\hat{Z}$ with parametrized fields $\{h^X(\lambda),h^Z(\lambda)\}$, where $\{\hat{X},\hat{Y},\hat{Z}\}$ are the Pauli matrices. 
The energy gap of this system is given by $\Delta=2\sqrt{h^{X}(\lambda)^2+h^{Z}(\lambda)^2}$ and the spectral function $\Phi(\omega)$ is nonzero only for $\omega=\pm\Delta$. 
Thus, the difference (\ref{Eq.distance}) becomes exactly zero for rotation angles $\theta_1=\pi/2\Delta$ and $\phi_1=-\delta\lambda/\Delta$ with $K=1$. 
For these rotation angles, the operator (\ref{Eq.AGP.ours}) is actually identical with the exact AGP of the two-level Hamiltonian~\cite{Demirplak2003,Demirplak2008,Berry2009} and the unitary operator (\ref{Eq.U}) approximates counterdiabatic driving (\ref{Eq.ad.trans}) as $\hat{U}(\lambda)=\hat{U}_\mathrm{ad}(\lambda)+\mathcal{O}(\delta\lambda^2)$ (see END MATTER for more details). 
Therefore, the basics of our method are consistent with the exact theory.

Next, we consider a many-body system $\hat{H}(\lambda)=[J(\lambda)/2N]\sum_{i,j=1}^N\hat{Z}_i\hat{Z}_j+h^X(\lambda)\sum_{i=1}^N\hat{X}_i$ with parameters $\{J(\lambda),h^X(\lambda)\}$, where $\{\hat{X}_i,\hat{Y}_i,\hat{Z}_i\}_{i=1}^N$ are the Pauli matrices of $N$ spins. 
The ground state of this system shows the second-order quantum phase transition at the critical point $|h^X(\lambda)/J(\lambda)|=1$ for $J(\lambda)<0$~\cite{Botet1982,Botet1983,Caneva2008,Yoshimura2014}. 
The system Hamiltonian commutes with the total-spin operator $\sum_{W=X,Y,Z}(\sum_{i=1}^N\hat{W_i})^2$ and the parity operator $\prod_{i=1}^N\hat{X}_i$, and thus dynamics is confined in $\lceil(N+1)/2\rceil$ dimension when the initial state is symmetric for these two operators. 
The ground state satisfies this condition.

\begin{figure}
\includegraphics[width=0.5\textwidth]{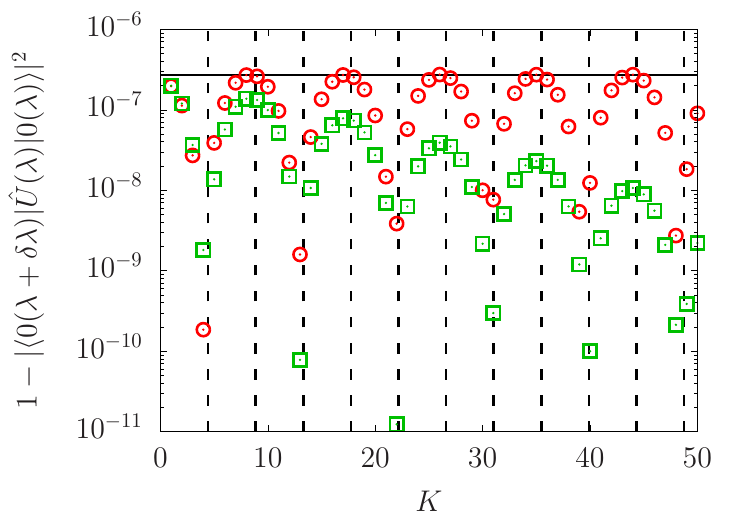}
\caption{\label{Fig.LMG}The infidelity to the ground state of the many-body system with respect to the integer $K$. The red circles are the results with the rotation angles (\ref{Eq.rot.angle}) and the green squares are those with the regularization for $\eta=0.1\Delta_\mathrm{min}$. The parameters are $J(\lambda)=J_0=-1$, $h^X(\lambda)=h_0^X\lambda=1$ ($h_0^X=1$ and $\lambda=1$), $\delta\lambda=10^{-3}$, $N=10$, and $\Omega=\Delta_\mathrm{max}=20.278\cdots$. The black solid line represents the infidelity under the parameter quench and the black dashed lines represent $kK_p^{\Omega=\Delta_\mathrm{max}}/2$ with $k=1,2,\dots$. }
\end{figure}

We numerically study parameter displacement of the ground state $|0(\lambda)\rangle$ at the critical point with $J(\lambda)=J_0=-1$, $h^X(\lambda)=h_0^X\lambda=1$ ($h_0^X=1$ and $\lambda=1$), $\delta\lambda=10^{-3}$, and $N=10$. 
For our method, we set $\Omega=\Delta_\mathrm{max}=20.278\cdots$. 
The infidelity to the ground state is plotted with respect to the integer $K$ in Fig.~\ref{Fig.LMG} [the red circles are the results with the rotation angles (\ref{Eq.rot.angle}) and the green squares are those with the regularization for $\eta=0.1\Delta_\mathrm{min}$]. 
We find the success in achieving very low infidelity in a periodic way with a period $K_p^{\Omega=\Delta_\mathrm{max}}=\Delta_\mathrm{max}/\Delta_\mathrm{min}$ (the black dashed lines represent $kK_p^{\Omega=\Delta_\mathrm{max}}/2$ with $k=1,2,\dots$), while the infidelity returns periodically (the black solid line represents the infidelity under a parameter quench, i.e., $1-|\langle0(\lambda+\delta\lambda)|0(\lambda)\rangle|^2$). 
Notably, the regularization can suppress the returns to the original value.

\begin{figure}
\includegraphics[width=0.5\textwidth]{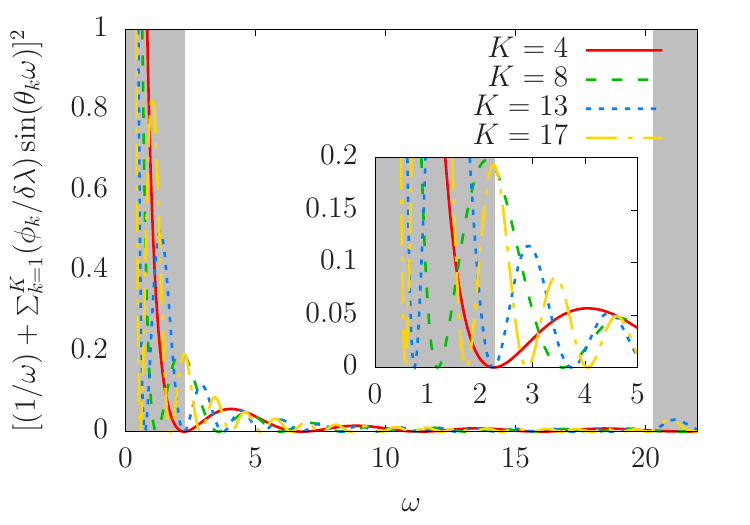}
\caption{\label{Fig.LMGkernel}The kernel that describes deviations from the exact AGP with respect to the energy differences. Each curve represents (red solid curve) $K=4$, (green dashed curve) $K=8$, (blue dotted curve) $K=13$, and (yellow dashed-dotted curve) $K=17$. The parameters are the same as Fig.~\ref{Fig.LMG}. The gray shaded areas are out of $[\Delta_\mathrm{min},\Delta_\mathrm{max}]$. The inset is an enlarged view around the minimum energy gap $\Delta_\mathrm{min}$. }
\end{figure}

We then analyze this periodic behavior with the rotation angles (\ref{Eq.rot.angle}). 
We focus on the kernel of the difference (\ref{Eq.distance}), i.e., $[(1/\omega)+\sum_{k=1}^K(\phi_k/\delta\lambda)\sin(\theta_k\omega)]^2$. 
It is known that the sine integral converges to $\pi/2$, i.e., $\lim_{x\to\infty}\mathrm{Si}(x)=\pi/2$, and accordingly $\phi_k$ also converges to $(-\pi\delta\lambda/\Omega)$ for large $k$. 
This means that the difference (\ref{Eq.distance}) does not converge to zero because the summand converges to $(-\pi/\Omega)\sin(k\pi\omega/\Omega)$, associated with the Gibbs phenomenon~\cite{Wilbraham1848,Gibbs1898,Gibbs1899} of the Fourier series due to the singularity of $(1/\omega)$ at $\omega=0$. 
By setting $k\to k^\prime K_p^{\Omega=\Delta_\mathrm{max}}/2$ ($k^\prime=1,2,\dots$) and $\omega=\Delta_\mathrm{min}$, the summand becomes $(-\pi/\Delta_\mathrm{max})\sin(k^\prime\pi/2)$. 
That is, when oscillation of the summand is canceled out, which is achieved for odd $k^\prime$, we can obtain very low infidelity. 
We also plot the kernel for several $K$ in Fig.~\ref{Fig.LMGkernel}. 
The gray shaded areas are out of $[\Delta_\mathrm{min},\Delta_\mathrm{max}]$, and thus it is important to suppress the kernel in the white area, particularly around the minimum energy gap. 
We actually find the cancellation of the summand for $K=4,13$ (corresponding to $k^\prime=1,3$) and accumulation of it for $K=8,17$ (corresponding to $k^\prime=2,4$). 
Note that the regularization removes the singularity of $(1/\omega)$ at $\omega=0$, and thus it resolves the Gibbs phenomenon.

\begin{figure}
\includegraphics[width=0.5\textwidth]{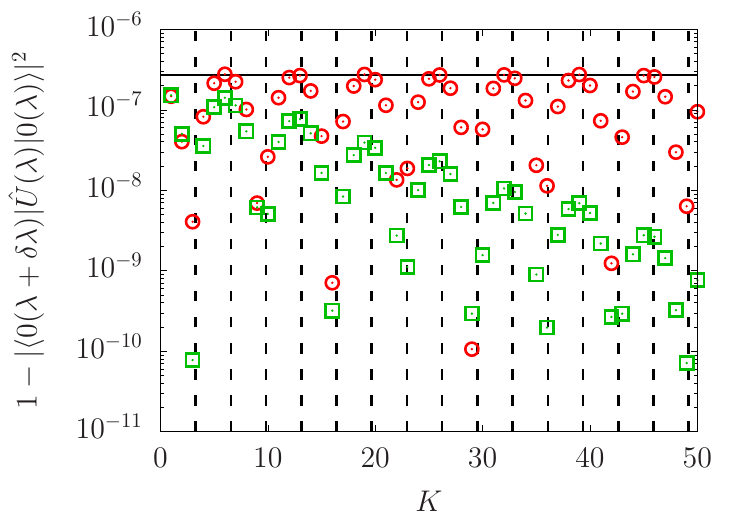}
\caption{\label{Fig.LMG.cutoff}The infidelity to the ground state of the many-body system with respect to the integer $K$. The symbols and parameters are the same as Fig.~\ref{Fig.LMG} except for the cutoff $\Omega=15$ instead of the maximum energy gap $\Delta_\mathrm{max}=20.278\cdots$. }
\end{figure}

Finally, we show that the cutoff $\Omega$ is not necessarily the maximum energy gap $\Delta_\mathrm{max}$ because the highest-energy eigenstate is irrelevant to the ground state. 
To exemplify this, we plot the infidelity to the ground state similarly to Fig.~\ref{Fig.LMG} except for the cutoff $\Omega=15$ instead of the maximum energy gap $\Delta_\mathrm{max}=20.278\cdots$ in Fig.~\ref{Fig.LMG.cutoff}. 
Remarkably, we find the achievement of very low infidelity and similar periodic behavior with a period $K_p^{\Omega=15}=15/\Delta_\mathrm{min}$ (the black dashed lines represent $kK_p^{\Omega=15}/2$ with $k=1,2,\dots$, and the black solid line is the same as Fig.~\ref{Fig.LMG}). 
Thus, we expect that the period of the oscillation is universally $K_p^\Omega=\Omega/\Delta_\mathrm{min}$ for a given cutoff $\Omega$. 
This expectation is also justified by the above analytical discussion.

%
%
{\it Discussion.---}
Now we compare our method with previously-proposed methods. 
The most relevant digital method was proposed in Ref.~\cite{Vreumingen2024} and benchmarked in Ref.~\cite{Hatomura2025}, which also used alternating unitary operators like Eq.~(\ref{Eq.U}). 
However, this method was derived by introducing several approximations to Eq.~(\ref{Eq.ad.trans}), and thus control fidelity was not very high and results were very noisy because of several approximation parameters. 
In contrast, our method has only two free parameters: the integer $K$ and the cutoff $\Omega$. 
Remarkably, our method has achieved very low infidelity, and the behavior of our method is periodic against the integer $K$ and robust against the cutoff $\Omega$. 
That is, our method is stable. 
Moreover, the complexity of our method is $\mathcal{O}(\Delta_\mathrm{min}^{-1})$ (see END MATTER for details), while that of their method is $\mathcal{O}(\Delta_\mathrm{min}^{-3})$.

More recently, another digital method was proposed in Ref.~\cite{Bhargava2026}. 
In this method, a unitary operator of the truncated AGP (\ref{Eq.AGP.nested}) is systematically decomposed into alternating unitary operators by using product formulae. 
A main disadvantage of this method is the requirement of the exponentially large number of unitary operators for composing nested commutators. 
In contrast, our method does not necessarily impose an exponential realization cost unless the minimum energy gap closes in an exponential way, while our method includes convincingly-controlled infinite nested commutators unlike their method. 
Moreover, our method gives the explicit expression of the rotation angles, while their method necessitates numerical calculation with the existing methods~\cite{Sels2017,Hatomura2021,Xie2022,Takahashi2024,Bhattacharjee2023,Ohga2025,Morawetz2025,Finzgar2025}.

Our method has used the idea of universal counterdiabatic driving~\cite{Morawetz2025,Finzgar2025}. 
Implementation of universal counterdiabatic driving is limited due to the difficulty in realizing nested commutators, which requires high-frequency driving or deep circuit depth. 
In our method, the smallest rotation angle, which corresponds to the inverse of the highest frequency, is $\phi_1=\mathcal{O}(\delta\lambda\Omega^{-1})$, and it can be adjusted because the result is robust against the change of the cutoff $\Omega$. 
The integer $K$, which is associated with the circuit depth, should be $K\approx K_p^\Omega/2=\Omega/2\Delta_\mathrm{min}$ to obtain very low infidelity, and it can also be adjusted simultaneously. 
Indeed, smaller $\Omega$ is more favorable for both quantities.

Our method has addressed the problems of the previous methods mentioned in the introduction: the unitary operator and the rotation angles are explicitly given without expensive computational cost; convincingly-controlled infinite nested commutators are generated, i.e., truncation is not necessary; and the requirement of high-frequency driving or deep circuit depth can be mitigated. 
A remaining limitation of our method is large $K$ required for the exponentially-small minimum energy gap $\Delta_\mathrm{min}$ as well as other methods.

%
%
{\it Conclusion and outlook.---}
In this paper, we have proposed universal digitized counterdiabatic driving, a universal method of constructing AGP and realizing counterdiabatic driving in a digital way. 
Our method does not require additional many-body and/or nonlocal interactions; includes convincingly-controlled infinite nested commutators; and gives the explicit expression of the rotation angles. 
We have shown that our method can achieve very low infidelity.

An open problem we leave is the way of determining optimal $K$ without explicit knowledge of the minimum energy gap. 
Another important direction extending the present work is the development of methods suppressing deviations around the minimum energy gap more efficiently and more stably. 
For example, standard approaches in the literature of the Fourier analysis, e.g., the use of a window function for suppressing the Gibbs phenomenon~\cite{Gottlieb1997}, are applicable. 
We expect that further improvement of our method could be possible by incorporating such techniques. 
In our demonstration, we just focus on the parameter displacement of the energy eigenstates, but AGP has various applications~\cite{Kato1950,Rigolin2008,Suzuki2020,Hatomura2020,You2007,Zanardi2007,Pandey2020,Lim2024} as mentioned in the introduction. 
Developing our method for those applications is the important future work to be addressed.

\begin{acknowledgments}
\section{Acknowledgment}
This work was supported by JST Moonshot R\&D Grant Number JPMJMS2061. 
The author is grateful to Jernej Rudi Fin\v{z}gar for their helpful comments and fruitful discussion. 
\end{acknowledgments}

\section{Data Availability}
The data that support the findings of this article are openly available~\cite{DVN/EWVCCL_2026}.

\bibliography{bib_UDCD}

@article{Albash2018,
   author = {Tameem Albash and Daniel A. Lidar},
   doi = {10.1103/RevModPhys.90.015002},
   issn = {0034-6861},
   issue = {1},
   journal = {Reviews of Modern Physics},
   month = {1},
   pages = {015002},
   publisher = {American Physical Society},
   title = {Adiabatic quantum computation},
   volume = {90},
   url = {https://link.aps.org/doi/10.1103/RevModPhys.90.015002},
   year = {2018},
}

@article{Berry2009,
   author = {M V Berry},
   doi = {10.1088/1751-8113/42/36/365303},
   issn = {1751-8113},
   issue = {36},
   journal = {Journal of Physics A: Mathematical and Theoretical},
   month = {9},
   pages = {365303},
   title = {Transitionless quantum driving},
   volume = {42},
   url = {http://stacks.iop.org/1751-8121/42/i=36/a=365303?key=crossref.8565bc95dce72711b5ffe1a34b22dc71},
   year = {2009},
}

@article{Bhargava2026,
   author = {Balaganchi A. Bhargava and Shubham Kumar and Anne-Maria Visuri and Paolo A. Erdman and Enrique Solano and Narendra N. Hegade},
   journal = {arXiv:2601.01154},
   month = {1},
   title = {{Constant Depth Digital-Analog Counterdiabatic Quantum Computing}},
   url = {https://arxiv.org/abs/2601.01154},
   year = {2026},
}

@article{Bhattacharjee2023,
   author = {Budhaditya Bhattacharjee},
   journal = {arXiv:2302.07228},
   month = {2},
   title = {{A Lanczos approach to the Adiabatic Gauge Potential}},
   url = {https://arxiv.org/abs/2302.07228},
   year = {2023},
}

@article{Born1928,
   author = {M Born and V Fock},
   doi = {10.1007/BF01343193},
   issn = {0044-3328},
   issue = {3},
   journal = {Zeitschrift für Physik},
   pages = {165-180},
   title = {{Beweis des Adiabatensatzes}},
   volume = {51},
   url = {https://doi.org/10.1007/BF01343193},
   year = {1928},
}

@article{Botet1982,
   author = {R. Botet and R. Jullien and P. Pfeuty},
   doi = {10.1103/PhysRevLett.49.478},
   issn = {0031-9007},
   issue = {7},
   journal = {Physical Review Letters},
   month = {8},
   pages = {478-481},
   publisher = {American Physical Society},
   title = {{Size Scaling for Infinitely Coordinated Systems}},
   volume = {49},
   url = {https://link.aps.org/doi/10.1103/PhysRevLett.49.478},
   year = {1982},
}

@article{Botet1983,
   author = {R. Botet and R. Jullien},
   doi = {10.1103/PhysRevB.28.3955},
   issn = {0163-1829},
   issue = {7},
   journal = {Physical Review B},
   month = {10},
   pages = {3955-3967},
   publisher = {American Physical Society},
   title = {Large-size critical behavior of infinitely coordinated systems},
   volume = {28},
   url = {https://link.aps.org/doi/10.1103/PhysRevB.28.3955},
   year = {1983},
}

@article{Caneva2008,
   author = {Tommaso Caneva and Rosario Fazio and Giuseppe E. Santoro},
   doi = {10.1103/PhysRevB.78.104426},
   issn = {1098-0121},
   issue = {10},
   journal = {Physical Review B},
   month = {9},
   pages = {104426},
   publisher = {American Physical Society},
   title = {{Adiabatic quantum dynamics of the Lipkin-Meshkov-Glick model}},
   volume = {78},
   url = {https://link.aps.org/doi/10.1103/PhysRevB.78.104426},
   year = {2008},
}

@article{Chen2022,
   author = {Yu-An Chen and Andrew M. Childs and Mohammad Hafezi and Zhang Jiang and Hwanmun Kim and Yijia Xu},
   doi = {10.1103/PhysRevResearch.4.013191},
   issn = {2643-1564},
   issue = {1},
   journal = {Physical Review Research},
   month = {3},
   pages = {013191},
   publisher = {American Physical Society},
   title = {Efficient product formulas for commutators and applications to quantum simulation},
   volume = {4},
   url = {https://link.aps.org/doi/10.1103/PhysRevResearch.4.013191},
   year = {2022},
}

@article{Claeys2019,
   author = {Pieter W. Claeys and Mohit Pandey and Dries Sels and Anatoli Polkovnikov},
   doi = {10.1103/PhysRevLett.123.090602},
   issn = {0031-9007},
   issue = {9},
   journal = {Physical Review Letters},
   month = {8},
   pages = {090602},
   publisher = {American Physical Society},
   title = {{Floquet-Engineering Counterdiabatic Protocols in Quantum Many-Body Systems}},
   volume = {123},
   url = {https://link.aps.org/doi/10.1103/PhysRevLett.123.090602},
   year = {2019},
}

@article{Demirplak2003,
   author = {Mustafa Demirplak and Stuart A Rice},
   doi = {10.1021/jp030708a},
   issn = {1089-5639},
   issue = {46},
   journal = {The Journal of Physical Chemistry A},
   month = {11},
   pages = {9937-9945},
   publisher = {American Chemical Society},
   title = {{Adiabatic Population Transfer with Control Fields}},
   volume = {107},
   url = {https://doi.org/10.1021/jp030708a},
   year = {2003},
}

@article{Demirplak2008,
   author = {Mustafa Demirplak and Stuart A. Rice},
   doi = {10.1063/1.2992152},
   issn = {0021-9606},
   issue = {15},
   journal = {The Journal of Chemical Physics},
   month = {10},
   pages = {154111},
   title = {On the consistency, extremal, and global properties of counterdiabatic fields},
   volume = {129},
   url = {http://aip.scitation.org/doi/10.1063/1.2992152},
   year = {2008},
}

@article{Finzgar2025,
   author = {Jernej Rudi Finžgar and Simone Notarnicola and Madelyn Cain and Mikhail D. Lukin and Dries Sels},
   doi = {10.1103/pqhl-nbtk},
   issn = {10797114},
   issue = {18},
   journal = {Physical Review Letters},
   month = {10},
   pages = {180602},
   pmid = {41247967},
   publisher = {American Physical Society},
   title = {{Counterdiabatic Driving with Performance Guarantees}},
   volume = {135},
   url = {https://journals.aps.org/prl/abstract/10.1103/pqhl-nbtk},
   year = {2025},
}

@article{Gibbs1898,
   author = {J. Willard Gibbs},
   doi = {10.1038/059200B0},
   issn = {1476-4687},
   issue = {1522},
   journal = {Nature},
   month = {12},
   pages = {200-200},
   publisher = {Nature Publishing Group},
   title = {{Fourier's Series}},
   volume = {59},
   url = {https://www.nature.com/articles/059200b0},
   year = {1898},
}

@article{Gibbs1899,
   author = {J. Willard Gibbs},
   doi = {10.1038/059606A0},
   issn = {00280836},
   issue = {1539},
   journal = {Nature},
   pages = {606},
   publisher = {Nature Publishing Group},
   title = {Fourier's series},
   volume = {59},
   url = {https://www.nature.com/articles/059606a0},
   year = {1899},
}

@article{Gottlieb1997,
   author = {David Gottlieb and Chi Wang Shu},
   doi = {10.1137/S0036144596301390},
   issn = {00361445},
   issue = {4},
   journal = {SIAM Review},
   month = {8},
   pages = {644-668},
   publisher = {Society for Industrial and Applied Mathematics},
   title = {{On the Gibbs Phenomenon and Its Resolution}},
   volume = {39},
   url = {/doi/pdf/10.1137/S0036144596301390?download=true},
   year = {1997},
}

@article{Guery-Odelin2019,
   author = {D. Gu\'ery-Odelin and A. Ruschhaupt and A. Kiely and E. Torrontegui and S. Mart\'inez-Garaot and J. G. Muga},
   doi = {10.1103/RevModPhys.91.045001},
   issn = {0034-6861},
   issue = {4},
   journal = {Reviews of Modern Physics},
   month = {10},
   pages = {045001},
   publisher = {American Physical Society},
   title = {{Shortcuts to adiabaticity: Concepts, methods, and applications}},
   volume = {91},
   url = {https://link.aps.org/doi/10.1103/RevModPhys.91.045001},
   year = {2019},
}

@article{Hatomura2020,
   author = {Takuya Hatomura and Go Kato},
   doi = {10.1103/PhysRevA.102.012216},
   issn = {2469-9926},
   issue = {1},
   journal = {Physical Review A},
   month = {7},
   pages = {012216},
   publisher = {American Physical Society},
   title = {Bounds for nonadiabatic transitions},
   volume = {102},
   url = {https://link.aps.org/doi/10.1103/PhysRevA.102.012216},
   year = {2020},
}

@article{Hatomura2021,
   author = {Takuya Hatomura and Kazutaka Takahashi},
   doi = {10.1103/PhysRevA.103.012220},
   issn = {2469-9926},
   issue = {1},
   journal = {Physical Review A},
   month = {1},
   pages = {012220},
   publisher = {American Physical Society},
   title = {Controlling and exploring quantum systems by algebraic expression of adiabatic gauge potential},
   volume = {103},
   url = {https://link.aps.org/doi/10.1103/PhysRevA.103.012220},
   year = {2021},
}

@article{Hatomura2023,
   author = {Takuya Hatomura},
   doi = {10.1088/1367-2630/acfd51},
   issn = {1367-2630},
   journal = {New Journal of Physics},
   month = {9},
   pages = {103025},
   title = {Scaling of errors in digitized counterdiabatic driving},
   volume = {25},
   url = {https://iopscience.iop.org/article/10.1088/1367-2630/acfd51},
   year = {2023},
}

@article{Hatomura2024,
   author = {Takuya Hatomura},
   doi = {10.1088/1361-6455/ad38f1},
   issn = {0953-4075},
   issue = {10},
   journal = {Journal of Physics B: Atomic, Molecular and Optical Physics},
   month = {3},
   pages = {102001},
   title = {Shortcuts to adiabaticity: theoretical framework, relations between different methods, and versatile approximations},
   volume = {57},
   url = {https://iopscience.iop.org/article/10.1088/1361-6455/ad38f1},
   year = {2024},
}

@article{Hatomura2025,
   author = {Takuya Hatomura},
   doi = {10.1103/PHYSREVA.111.022411},
   issn = {24699934},
   issue = {2},
   journal = {Physical Review A},
   month = {2},
   pages = {022411},
   publisher = {American Physical Society},
   title = {Benchmarking adiabatic transformation by alternating unitaries},
   volume = {111},
   url = {https://journals.aps.org/pra/abstract/10.1103/PhysRevA.111.022411},
   year = {2025},
}

@article{Hegade2021,
   author = {Narendra N. Hegade and Koushik Paul and Yongcheng Ding and Mikel Sanz and F. Albarrán-Arriagada and Enrique Solano and Xi Chen},
   doi = {10.1103/PhysRevApplied.15.024038},
   issn = {2331-7019},
   issue = {2},
   journal = {Physical Review Applied},
   month = {2},
   pages = {024038},
   publisher = {American Physical Society},
   title = {{Shortcuts to Adiabaticity in Digitized Adiabatic Quantum Computing}},
   volume = {15},
   url = {https://link.aps.org/doi/10.1103/PhysRevApplied.15.024038},
   year = {2021},
}

@article{Hegade2022,
   author = {Narendra N. Hegade and Xi Chen and Enrique Solano},
   doi = {10.1103/PhysRevResearch.4.L042030},
   issn = {2643-1564},
   issue = {4},
   journal = {Physical Review Research},
   month = {11},
   pages = {L042030},
   publisher = {American Physical Society},
   title = {Digitized counterdiabatic quantum optimization},
   volume = {4},
   url = {https://link.aps.org/doi/10.1103/PhysRevResearch.4.L042030},
   year = {2022},
}

@article{Kato1950,
   author = {Tosio Kato},
   doi = {10.1143/JPSJ.5.435},
   issn = {0031-9015},
   issue = {6},
   journal = {Journal of the Physical Society of Japan},
   month = {11},
   pages = {435-439},
   publisher = {The Physical Society of Japan},
   title = {{On the Adiabatic Theorem of Quantum Mechanics}},
   volume = {5},
   url = {http://journals.jps.jp/doi/10.1143/JPSJ.5.435},
   year = {1950},
}

@article{Kolodrubetz2017,
   author = {Michael Kolodrubetz and Dries Sels and Pankaj Mehta and Anatoli Polkovnikov},
   doi = {10.1016/J.PHYSREP.2017.07.001},
   issn = {0370-1573},
   journal = {Physics Reports},
   month = {6},
   pages = {1-87},
   publisher = {North-Holland},
   title = {Geometry and non-adiabatic response in quantum and classical systems},
   volume = {697},
   url = {https://www.sciencedirect.com/science/article/abs/pii/S0370157317301989},
   year = {2017},
}

@article{Lim2024,
   author = {Cedric Lim and Kirill Matirko and Anatoli Polkovnikov and Michael O. Flynn},
   journal = {arXiv:2401.01927},
   month = {1},
   title = {Defining classical and quantum chaos through adiabatic transformations},
   url = {https://doi.org/10.48550/arXiv.2401.01927},
   year = {2024},
}

@article{Keever2024,
   author = {Conor Mc Keever and Michael Lubasch},
   doi = {10.1103/PRXQuantum.5.020362},
   issn = {2691-3399},
   issue = {2},
   journal = {PRX Quantum},
   month = {6},
   pages = {020362},
   publisher = {American Physical Society},
   title = {{Towards Adiabatic Quantum Computing Using Compressed Quantum Circuits}},
   volume = {5},
   url = {https://link.aps.org/doi/10.1103/PRXQuantum.5.020362},
   year = {2024},
}

@article{Morawetz2025,
   author = {Stewart Morawetz and Anatoli Polkovnikov},
   doi = {10.1103/wbbs-s8fs},
   issue = {4},
   journal = {PRX Quantum},
   month = {10},
   pages = {040320},
   publisher = {American Physical Society},
   title = {{Universal Counterdiabatic Driving in Krylov Space}},
   volume = {6},
   url = {https://journals.aps.org/prxquantum/abstract/10.1103/wbbs-s8fs},
   year = {2025},
}

@article{Ohga2025,
   author = {Naruo Ohga and Takuya Hatomura},
   journal = {arXiv:2505.18367},
   month = {5},
   title = {Improving variational counterdiabatic driving with weighted actions and computer algebra},
   url = {https://doi.org/10.48550/arXiv.2505.18367},
   year = {2025},
}

@article{Pandey2020,
   author = {Mohit Pandey and Pieter W. Claeys and David K. Campbell and Anatoli Polkovnikov and Dries Sels},
   doi = {10.1103/PhysRevX.10.041017},
   issn = {2160-3308},
   issue = {4},
   journal = {Physical Review X},
   month = {10},
   pages = {041017},
   publisher = {American Physical Society},
   title = {{Adiabatic Eigenstate Deformations as a Sensitive Probe for Quantum Chaos}},
   volume = {10},
   url = {https://link.aps.org/doi/10.1103/PhysRevX.10.041017},
   year = {2020},
}

@article{Petiziol2018,
   author = {Francesco Petiziol and Benjamin Dive and Florian Mintert and Sandro Wimberger},
   doi = {10.1103/PhysRevA.98.043436},
   issn = {2469-9926},
   issue = {4},
   journal = {Physical Review A},
   month = {10},
   pages = {043436},
   publisher = {American Physical Society},
   title = {{Fast adiabatic evolution by oscillating initial Hamiltonians}},
   volume = {98},
   url = {https://link.aps.org/doi/10.1103/PhysRevA.98.043436},
   year = {2018},
}

@article{Rigolin2008,
   author = {Gustavo Rigolin and Gerardo Ortiz and Victor Hugo Ponce},
   doi = {10.1103/PhysRevA.78.052508},
   issn = {10502947},
   issue = {5},
   journal = {Physical Review A},
   month = {11},
   pages = {052508},
   publisher = {American Physical Society},
   title = {{Beyond the quantum adiabatic approximation: Adiabatic perturbation theory}},
   volume = {78},
   url = {https://journals.aps.org/pra/abstract/10.1103/PhysRevA.78.052508},
   year = {2008},
}

@article{Sels2017,
   author = {Dries Sels and Anatoli Polkovnikov},
   doi = {10.1073/pnas.1619826114},
   issn = {1091-6490},
   issue = {20},
   journal = {Proceedings of the National Academy of Sciences of the United States of America},
   month = {5},
   pages = {E3909-E3916},
   pmid = {28461472},
   publisher = {National Academy of Sciences},
   title = {Minimizing irreversible losses in quantum systems by local counterdiabatic driving.},
   volume = {114},
   url = {http://www.ncbi.nlm.nih.gov/pubmed/28461472 http://www.pubmedcentral.nih.gov/articlerender.fcgi?artid=PMC5441767},
   year = {2017},
}

@article{Suzuki2020,
   author = {Keisuke Suzuki and Kazutaka Takahashi},
   doi = {10.1103/PhysRevResearch.2.032016},
   issn = {2643-1564},
   issue = {3},
   journal = {Physical Review Research},
   month = {7},
   pages = {032016(R)},
   publisher = {American Physical Society},
   title = {Performance evaluation of adiabatic quantum computation via quantum speed limits and possible applications to many-body systems},
   volume = {2},
   url = {https://link.aps.org/doi/10.1103/PhysRevResearch.2.032016},
   year = {2020},
}

@article{Takahashi2024,
   author = {Kazutaka Takahashi and Adolfo del Campo},
   doi = {10.1103/PhysRevX.14.011032},
   issn = {2160-3308},
   issue = {1},
   journal = {Physical Review X},
   month = {2},
   pages = {011032},
   publisher = {American Physical Society},
   title = {{Shortcuts to Adiabaticity in Krylov Space}},
   volume = {14},
   url = {https://link.aps.org/doi/10.1103/PhysRevX.14.011032},
   year = {2024},
}

@article{Torrontegui2013,
   author = {Erik Torrontegui and Sara Ibáñez and Sofia Martínez-Garaot and Michele Modugno and Adolfo del Campo and David Guéry-Odelin and Andreas Ruschhaupt and Xi Chen and Juan Gonzalo Muga},
   doi = {10.1016/B978-0-12-408090-4.00002-5},
   isbn = {9780124080904},
   issn = {1049-250X},
   journal = {Advances In Atomic, Molecular, and Optical Physics},
   month = {1},
   pages = {117-169},
   publisher = {Academic Press},
   title = {{Shortcuts to Adiabaticity}},
   volume = {62},
   url = {https://www.sciencedirect.com/science/article/pii/B9780124080904000025?via%3Dihub},
   year = {2013},
}

@article{Vizzuso2025,
   author = {Mara Vizzuso and Gianluca Passarelli and Giovanni Cantele and Procolo Lucignano and Xi Chen and Koushik Paul},
   journal = {arXiv:2512.22636},
   month = {12},
   title = {{Nonadiabatic Self-Healing of Trotter Errors in Digitized Counterdiabatic Dynamics}},
   url = {https://arxiv.org/abs/2512.22636},
   year = {2025},
}

@article{Vreumingen2024,
   author = {Dyon van Vreumingen},
   doi = {10.1103/PhysRevA.110.052419},
   issn = {2469-9926},
   issue = {5},
   journal = {Physical Review A},
   month = {11},
   pages = {052419},
   publisher = {American Physical Society},
   title = {Gate-based counterdiabatic driving with complexity guarantees},
   volume = {110},
   url = {https://journals.aps.org/pra/abstract/10.1103/PhysRevA.110.052419},
   year = {2024},
}

@article{Wilbraham1848,
   author = {Henry Wilbraham},
   journal = {The Cambridge and Dublin Mathematical Journal},
   pages = {198-201},
   title = {On a certain periodic function},
   volume = {3},
   year = {1848},
}

@article{Xie2022,
   author = {Qing Xie and Kazuhiro Seki and Seiji Yunoki},
   doi = {10.1103/PhysRevB.106.155153},
   issn = {2469-9950},
   issue = {15},
   journal = {Physical Review B},
   month = {10},
   pages = {155153},
   publisher = {American Physical Society},
   title = {{Variational counterdiabatic driving of the Hubbard model for ground-state preparation}},
   volume = {106},
   url = {https://link.aps.org/doi/10.1103/PhysRevB.106.155153},
   year = {2022},
}

@article{Yoshimura2014,
   author = {Bryce Yoshimura and W. C. Campbell and J. K. Freericks},
   doi = {10.1103/PhysRevA.90.062334},
   issn = {1050-2947},
   issue = {6},
   journal = {Physical Review A},
   month = {12},
   pages = {062334},
   publisher = {American Physical Society},
   title = {Diabatic-ramping spectroscopy of many-body excited states},
   volume = {90},
   url = {https://link.aps.org/doi/10.1103/PhysRevA.90.062334},
   year = {2014},
}

@article{You2007,
   author = {Wen Long You and Ying Wai Li and Shi Jian Gu},
   doi = {10.1103/PhysRevE.76.022101},
   issn = {15393755},
   issue = {2},
   journal = {Physical Review E},
   month = {8},
   pages = {022101},
   publisher = {American Physical Society},
   title = {Fidelity, dynamic structure factor, and susceptibility in critical phenomena},
   volume = {76},
   url = {https://journals.aps.org/pre/abstract/10.1103/PhysRevE.76.022101},
   year = {2007},
}

@article{Zanardi2007,
   author = {Paolo Zanardi and Paolo Giorda and Marco Cozzini},
   doi = {10.1103/PhysRevLett.99.100603},
   issn = {0031-9007},
   issue = {10},
   journal = {Physical Review Letters},
   month = {9},
   pages = {100603},
   publisher = {American Physical Society},
   title = {{Information-Theoretic Differential Geometry of Quantum Phase Transitions}},
   volume = {99},
   url = {https://link.aps.org/doi/10.1103/PhysRevLett.99.100603},
   year = {2007},
}

@article{DVN/EWVCCL_2026,
author = {Hatomura, Takuya},
journal = {Harvard Dataverse},
title = {{Replication Data for: Universal Digitized Counterdiabatic Driving}},
year = {2026},
version = {V1},
doi = {10.7910/DVN/EWVCCL},
url = {https://doi.org/10.7910/DVN/EWVCCL}
}

\section*{END MATTER}
\subsection*{Details for the general theory of universal digitized counterdiabatic driving}
First, we derive Eq.~(\ref{Eq.AGP.ours}). 
Equation (\ref{Eq.U}) can be rewritten as
\begin{equation}
\begin{aligned}
\hat{U}(\lambda)=&\prod_{\substack{k=-K \\ (k\neq0)}}^Ke^{-i\frac{\phi_k}{2}e^{i\theta_k\hat{H}(\lambda)}\partial_\lambda\hat{H}(\lambda)e^{-i\theta_k\hat{H}(\lambda)}} \\
=&\prod_{\substack{k=-K \\ (k\neq0)}}^Ke^{-i\frac{\phi_k}{2}\sum_{l=0}^\infty\frac{(i\theta_k)^l}{l!}\mathcal{L}^l\partial_\lambda\hat{H}(\lambda)} \\
=&e^{-i\sum_{k=-K (k\neq0)}^K\frac{\phi_k}{2}\sum_{l=0}^\infty\frac{(i\theta_k)^l}{l!}\mathcal{L}^l\partial_\lambda\hat{H}(\lambda)}+\mathcal{O}(\{\phi_k^2\}),
\end{aligned}
\label{Eq.der.digU}
\end{equation}
where we use a property of unitary operators $(e^{i\hat{A}}e^{i\hat{B}}e^{-i\hat{A}}=e^{ie^{i\hat{A}}\hat{B}e^{-i\hat{A}}})$ in the first line, the Baker-Campbell-Hausdorff formula in the second line, and the Lie-Trotter formula in the third line. 
Application of the Lie-Trotter formula requires $|\phi_k|\ll1$ to suppress the Trotter error. 
Then, the conditions, $\theta_k=-\theta_{-k}$ and $\phi_k=-\phi_{-k}$, result in Eq.~(\ref{Eq.AGP.ours}).

Next, we derive Eq.~(\ref{Eq.distance}). 
The Hilbert-Schmidt norm is given by the sum of the absolute squares of the all matrix elements. 
The matrix element of AGP (\ref{Eq.AGP}) is given by $\langle m(\lambda)|\hat{\mathcal{A}}(\lambda)|n(\lambda)\rangle=i\langle m(\lambda)|\partial_\lambda n(\lambda)\rangle=i\langle m(\lambda)|\partial_\lambda\hat{H}(\lambda)|n(\lambda)\rangle/\omega_{nm}$ and that of the nested commutator $\mathcal{L}^{2l-1}\partial_\lambda\hat{H}(\lambda)$ is given by $\langle m(\lambda)|\mathcal{L}^{2l-1}\partial_\lambda\hat{H}(\lambda)|n(\lambda)\rangle=\omega_{mn}^{2l-1}\langle m(\lambda)|\partial_\lambda\hat{H}(\lambda)|n(\lambda)\rangle$. 
That is, the difference between the exact AGP (\ref{Eq.AGP}) and the operator (\ref{Eq.AGP.ours}) can be calculated as
\begin{equation}
\begin{aligned}
&\|\hat{\mathcal{A}}(\lambda)-\hat{V}(\lambda)\|^2=\\
&\sum_{\substack{m,n=0 \\ (m\neq n)}}^{D-1}\left[\frac{1}{\omega_{mn}}+\sum_{k=1}^K\frac{\phi_k}{\delta\lambda}\sin(\theta_k\omega_{mn})\right]^2|\langle m(\lambda)|\partial\hat{H}(\lambda)|n(\lambda)\rangle|^2,
\end{aligned}
\end{equation}
which is equivalent to Eq.~(\ref{Eq.distance}). 
Note that the diagonal elements of the AGP (\ref{Eq.AGP}) and the operator  (\ref{Eq.AGP.ours}) are zeros.

Finally, we consider the specialization for the ground state. 
For this purpose, we use the difference $\|[\hat{\mathcal{A}}(\lambda)-\hat{V}(\lambda)]|0(\lambda)\rangle\|^2$ instead of Eq.~(\ref{Eq.distance}), where we adopt the Euclidean norm and $|0(\lambda)\rangle$ is the ground state. 
Then we obtain the right-hand side of Eq.~(\ref{Eq.distance}) with the spectral function $\Phi(\omega)=\sum_{m=1}^{D-1}|\langle m(\lambda)|\partial_\lambda\hat{H}(\lambda)|0(\lambda)\rangle|^2\delta(\omega\pm\omega_{m0})$. 
Thus, the minimum energy gap can be specialized for the ground state.

\subsection{Details for the case study of the two-level system}
We show that our method can generate the exact AGP of the two-level system $\hat{H}(\lambda)=h^X(\lambda)\hat{X}+h^Z(\lambda)\hat{Z}$, 
\begin{equation}
\hat{\mathcal{A}}(\lambda)=\frac{h^Z(\lambda)\partial_\lambda h^X(\lambda)-h^X(\lambda)\partial_\lambda h^Z(\lambda)}{2[h^X(\lambda)^2+h^Z(\lambda)^2]}\hat{Y}, 
\end{equation}
with the rotation angles $\theta_1=\pi/2\Delta$ and $\phi_1=-\delta\lambda/\Delta$ for $K=1$. 
The odd nested commutator is calculated as
\begin{equation}
\mathcal{L}^{2l-1}\partial_\lambda\hat{H}(\lambda)=2i[h^Z(\lambda)\partial_\lambda h^X(\lambda)-h^X(\lambda)\partial_\lambda h^Z(\lambda)]\Delta^{2l-2}\hat{Y}. 
\end{equation}
By substituting this and the rotation angles for Eq.~(\ref{Eq.AGP.ours}), we immediately find $\hat{V}(\lambda)=\hat{\mathcal{A}}(\lambda)$. 
Then Eq.~(\ref{Eq.der.digU}) gives $\hat{U}(\lambda)=\hat{U}_\mathrm{ad}(\lambda)+\mathcal{O}(\phi_1^2)=\hat{U}_\mathrm{ad}(\lambda)+\mathcal{O}(\delta\lambda^2)$, which supports the main text.

\subsection{Complexity of universal digitized counterdiabatic driving}
For a given digital, composite unitary operator, $\prod_ke^{i\hat{h}_kt_k}$, we define its complexity as $\sum_k\|\hat{h}_k\||t_k|$ with given norm. 
Then, considering the cancellation of unitary operators $e^{-i\theta_k\hat{H}(\lambda)}e^{i\theta_{k+1}\hat{H}(\lambda)}=e^{i\frac{\pi}{\Omega}\hat{H}(\lambda)}$, the complexity of our method is evaluated as
\begin{equation}
\frac{4K\pi}{\Omega}\|\hat{H}(\lambda)\|+\frac{2\delta\lambda}{\Omega}\sum_{k=1}^K\mathrm{Si}(k\pi)\|\partial_\lambda\hat{H}(\lambda)\|. 
\end{equation}
We expect that the first term is dominant because $\mathrm{Si}(x)=\mathcal{O}(1)$ and $\delta\lambda\|\partial_\lambda\hat{H}(\lambda)\|\approx\|\hat{H}(\lambda+\delta\lambda)-\hat{H}(\lambda)\|$ should be small. 
To obtain very low infidelity, $K$ should be $K_p^\Omega/2=\Omega/2\Delta_\mathrm{min}$, and thus the complexity of universal digitized counterdiabatic driving scales as $\mathcal{O}(\|\hat{H}(\lambda)\|\Delta_\mathrm{min}^{-1})$.

\end{document}